# Electric field-induced superconducting transition of insulating FeSe thin film at 35 K

**Authors:** Kota Hanzawa,[1] Hikaru Sato,[1] Hidenori Hiramatsu,[1,2*] Toshio Kamiya,[1,2] Hideo Hosono,[1,2*]

**Affiliations:**

[1]Materials and Structures Laboratory, Tokyo Institute of Technology, Mailbox R3-1, 4259 Nagatsuta-cho, Midori-ku, Yokohama 226-8503, Japan.

[2]Materials Research Center for Element Strategy, Tokyo Institute of Technology, Mailbox SE-6, 4259 Nagatsuta-cho, Midori-ku, Yokohama 226-8503, Japan.

*Correspondence to: h-hirama@lucid.msl.titech.ac.jp; hosono@msl.titech.ac.jp

**Abstract**

It is thought that strong electron correlation in an insulating parent phase would enhance a critical temperature ($T_c$) of superconductivity in a doped phase via enhancement of the binding energy of a Cooper pair as known in high-$T_c$ cuprates. To induce a superconductor transition in an insulating phase, injection of a high density of carriers is needed (e.g., by impurity doping). An electric double-layer transistor (EDLT) with an ionic liquid gate insulator enables such a field-induced transition to be investigated and is expected to result in a high $T_c$ because it is free from deterioration in structure and carrier transport that are in general caused by conventional carrier doping (e.g., chemical substitution). Here, for insulating epitaxial thin films (~10 nm thick) of FeSe, we report a high $T_c$ of 35 K, which is four times higher than that of bulk FeSe, using an EDLT under application of a gate bias of +5.5 V. Hall effect measurements under the gate bias suggest that highly accumulated electron carrier in the channel, whose area density is estimated to be $1.4\times10^{15}$ cm$^{-2}$ (the average volume density of $1.7\times10^{21}$ cm$^{-3}$), is the origin of the high-$T_c$ superconductivity. This result demonstrates that EDLTs are useful tools to explore the ultimate $T_c$ for insulating parent materials.

Prediction of new high critical temperature ($T_c$) superconductors is one of the most difficult issues in condensed matter physics, although theories and calculation methods have recently been very rapidly advancing (1). There is a common feature of high-$T_c$ superconductors, copper-based oxides (cuprates) (2), and iron-based pnictides/selenides (3): Their superconductivity emerges when the long-range antiferromagnetic order in their parent phases is suppressed by doping with a high-density of carriers. However, the maximum $T_c$ of iron-based pnictides/selenides [55 K for $SmFeAs(O_{1-x}F_x)$] (4) is much lower than that of cuprates (134 K for $HgBa_2Ca_2Cu_3O_{8+\delta}$) (5), although a very thin monolayer of FeSe with $T_c$ = 109 K has been reported exceptionally (6). This difference in $T_c$ is related to the different electron correlation interactions in these material systems. The parent phase of cuprates is a Mott insulator in which electron–electron Coulomb repulsion is very strong, whereas almost all of the iron-based parent phases are "(poor) metals", not insulators, because of weak electron correlations. According to this high-$T_c$ cuprate scenario, we consider that an important strategy to obtain higher $T_c$ in iron-based superconductors is to dope carriers into the insulating parent phase with strong electron correlation to induce an insulator–superconductor transition.

In 2014, we first reported an electric field-induced phase transition in the iron-based insulator $TlFe_{1.6}Se_2$ (7), to the best of our knowledge. An electric double-layer transistor (EDLT) structure with an ion-liquid gate was used to examine electrostatic carrier doping because the ion-liquid gate has a large capacitance and provides an effective way to accumulate carriers at much higher densities (e.g., maximum sheet carrier density of ~$10^{15}$ cm$^{-2}$) (8–11) than those obtained by conventional solid-state gate materials such as amorphous $SiO_x$ ($10^{12}$–$10^{13}$ cm$^{-2}$). The $TlFe_{1.6}Se_2$ EDLT showed that using EDLTs is an effective way to explore phase transitions in iron-based high-$T_c$ pnictides/selenides, where carrier doping by conventional chemical substitution is difficult. A phase transition from a superconducting state to an insulating state has been observed in an EDLT using single crystals of (Li,Fe)OHFeSe (12). In addition, an enhancement of $T_c$ has been observed in EDLTs using superconducting FeSe (13). However, direct induction of superconductivity from an insulating phase has not been achieved in EDLTs using iron-based layered pnictides/selenides, although this type of phase transition is very drastic and interesting.

Very thin (~10-nm-thick) FeSe epitaxial films are an interesting material to investigate field effects and to explore high $T_c$ because thin FeSe films exhibit insulator-like behavior (14, 15). Considering the fact that bulk FeSe is a superconductor with $T_c \sim 8$ K (16), the origin of the insulator-like behavior in the FeSe epitaxial films is thought to originate from lattice strain (15), which would be similar to a recently reported strain-induced superconductivity in $BaFe_2As_2$ (17). Furthermore, it has been reported that monolayer-thick FeSe exhibits very high $T_c$ values of 50–109 K (6, 18). These results imply that an epitaxial strain has a key factor to controlling its electronic phase diagram in iron-based superconductors.

In this study, we focused on EDLT-based field effects of ~10-nm-thick FeSe epitaxial films and observed an insulator–superconductor transition with a highest $T_c$ of 35 K under a gate bias of $V_G = +5.5$ V. The origin of this superconductivity is also discussed.

Fig. 1 summarizes the structure and the relationship between the temperature ($T$) and the resistivity ($\rho$) of ~10-nm-thick FeSe epitaxial films on $SrTiO_3$ (STO) (001) grown by molecular beam epitaxy (MBE). Fig. 1A and 1B shows an out-of-plane X-ray diffraction (XRD) pattern and an out-of-plane X-ray rocking curve (XRC) of FeSe 001 diffraction, respectively, indicating strong $c$-axis preferential orientation of the films (a wide-range pattern for $2\theta = 10$–$80°$ is shown in Fig. S1A, confirming that there is no impurity phase). Because of the atomically flat surface of STO substrate, clear Kikuchi patterns in the reflective high-energy electron diffraction (RHEED) pattern, a step-and-terrace structure with a terrace width of ~120 nm, and no step-bunching are observed in Fig. 1C and 1E (see the cross-section of the STO substrate surface in Fig. S2B). Heteroepitaxial growth of the FeSe film on STO is confirmed by the RHEED pattern in Fig. 1D ($2\theta_\chi$-fixed $\phi$ scan of FeSe 200 in-plane diffraction is also shown in Fig. S1B), and the film has an atomically flat surface with root-mean-square roughness of 0.6 nm (Fig. 1F). It was confirmed that the chemical composition is 1.1:1 in the Fe:Se atomic ratio (measured with an electron probe microanalyzer). Fig. 1G shows $\rho$–$T$ curves of the 11-nm-thick FeSe film compared with a 110-nm-thick film. The thick film (110 nm) did not show zero resistance when $T$ was decreased to 4 K, whereas superconductivity with an onset $T_c \sim 9$ K was confirmed by its external magnetic field dependence (Fig. 1H), i.e., a negative shift of $T_c$ with increasing external magnetic field. However, in the case of the thin films (11 nm thick, see Fig. S3 for determination of the thickness by X-ray reflectivity measurement), the $\rho$–$T$ curve shows insulator-like behavior, i.e., $\rho$ increased with decreasing $T$. The lattice parameters of the 11-nm-thick films are $a_{\text{film}} = 0.3838$ nm for the $a$ axis (see Fig. S4 for in-plane XRD patterns) and $c_{\text{film}} = 0.5448$ nm for the $c$ axis (calculated from the peak position of out-of-plane FeSe 004 diffraction, Fig. S1A). Because the values of a bulk sample are reported to be $a_{\text{bulk}} = 0.37704$ nm and $c_{\text{bulk}} = 0.55161$ nm (19), there is in-plane tensile strain [$(a_{\text{film}}-a_{\text{bulk}})/a_{\text{bulk}} = \Delta a/a_{\text{bulk}} = +1.8\%$] but out-of-plane compressive strain ($\Delta c/c_{\text{bulk}} = -1.2\%$). These results indicate that the insulator-like behavior of the epitaxial film originates from the in-plane tensile strain, as reported in ref. 15. Because the insulating state is the main focus of this study, we selected the thin (~10 nm) FeSe epitaxial layer as an active channel layer for the EDLT.

We fabricated EDLTs using a thin insulating FeSe layer, as shown in Fig. 2A. The ionic liquid $N$,$N$-diethyl-$N$-methyl-$N$-(2-methoxyethyl)-ammonium bis-(trifluoromethylsulfonyl)imide (DEME-TFSI) was used as the gate insulator. Fig. 2B shows the cyclic transfer characteristics [drain current ($I_D$) versus gate bias ($V_G$)] of the FeSe EDLT under a drain voltage $V_D = +0.5$ V measured at $T = 220$ K. A positive $V_G$ up to +4 V was applied to the gate electrode, which accumulates carrier electrons in the FeSe surface. When $V_G = +3.1$ V was applied, $I_D$ began increasing. The maximum $I_D$ in the transfer curve reached 45 μA at $V_G = +4$ V, along with an on–off ratio of ~2. The gate leakage current ($I_G$, shown in Fig. 2B, Bottom) also increased with increasing $V_G$ up to +4 V, but it was three orders of magnitude lower than $I_D$ in the whole $V_G$ region. After applying $V_G = +4$ V, $I_D$ returned to the initial value when $V_G$ was decreased to 0 V. A large hysteresis loop is observed because of the slow response of ion displacement in the ionic liquid. Notwithstanding there is a slight parallel shift in the second $I_D$−$V_G$ loop and $I_D$ becomes almost double, 84 μA at $V_G = +4$ V; its origin is considered to be the slow response of ion displacement, which is the same as the hysteresis loop. This is because the shape and hysteresis width in the second loop are very similar to those in the first loop. In addition, we confirmed that there was no change in the $I$–$V$ characteristics in the channel layer at $V_G = 0$ before and after the cyclic measurements (see Fig. S5 for the $I$–$V$ characteristics). Here, we point out that it has been reported that EDLT using an STO channel exhibits similar characteristics, i.e., it also turns on at about $V_G = +3$ V (8). In addition, it has recently been reported that oxygen vacancies are induced in an oxide-based EDLT and the metal state is stabilized even after removing the gate bias and ionic liquid (10). However, the above results in Fig. 2 and Fig. S5 guarantee that the observed results are reversible and reproducible. These results demonstrate that the origin of superconductivity by applying gate bias, which will be shown in Fig. 3 later, is not due to the modification of the surface chemical/mechanical structures but the electrostatically accumulated carriers.

Fig. 3 summarizes carrier transport properties of the FeSe EDLT, where $V_G$ was applied at 220 K and kept constant during the measurements with decreasing $T$. As shown in Fig. 3A, the sheet resistance ($R_s$)–$T$ curves at $V_G = 0 - +3.5$ V almost overlap in the whole $T$ range. In contrast, when $V_G$ is increased to +3.75 V, $R_s$ in the normal-state region slightly decreases, indicating that the induced carrier density is increased by applying $V_G = +3.75$ V. When $V_G = +4$ V is applied, a broad $R_s$ drop is observed at 8.6 K (see Fig. 3B). With further increase in $V_G$, this $R_s$ drop shifts to higher $T$. It should be noted that zero resistance is clearly observed at 4 K along with the onset $T = 24$ K when $V_G = +5$ V is applied. From $V_G = +5$ V to +5.5 V, there is a clear enhancement of the onset $T$ of the $R_s$ drop (from 24 to 35 K) along with reduction of the normal-state resistivity. Fig. 3C shows the magnetic field dependence at $V_G = +5.5$ V, confirming its superconducting transition, i.e., the onset $T_c$ shifts to lower $T$ with increasing external magnetic field. The superconducting state remains even when the external magnetic field is increased to 9 T, indicating that the upper critical magnetic field is considerably higher than 9 T. The upper critical field estimated from Fig. 3C using Werthamer–Helfand–Hohenberg theory (20) is ~70 T (Fig. S6), which is much higher than that (30 T) of bulk FeSe (21).

To confirm high-density carrier accumulation in the EDLT, we performed Hall effect measurements under applying $V_G$ up to +5.5 V. Fig. 3D shows Hall coefficients $R_H$ at 40 K. Positive $R_H$ are obtained at each $V_G$ applied, indicating that major carrier in the insulator-like FeSe epitaxial films is hole and consistent with the result in an FeSe single crystal (22), although negative $R_H$ in low magnetic fields is also reported in this temperature range (23). At $V_G = 0$ V, the $R_H$ is $+1.1 \times 10^{-2}$ cm$^3$/C, which is slightly higher than ($+2 \times 10^{-3}$ cm$^3$/C) as reported in ref. 22. With increase in $V_G$ to +3.5 V, $R_H$ remains almost constant, which is consistent with $R_s$–$T$ curves in Fig. 3A; i.e., they almost overlap in the whole $T$ range. With further increase in $V_G$ up to +5.5 V, $R_H$ decreases linearly to $+2.7 \times 10^{-3}$ cm$^3$/C. It is reported that FeSe bulk is a multiband metal, indicating that high-density electrons and holes (both are orders of $10^{20}$ cm$^{-3}$.) (23) intrinsically coexist and make its carrier transport properties very complicated. In addition, the electronic structure of this insulator-like FeSe epitaxial film is still unclear. However, to roughly estimate the electron density accumulated in the channel surface by positive $V_G$, we subtracted the $1 / eR_H$ value (corresponds to the carrier density for a single band model) at $V_G = 0$ V from all the $1 / eR_H$ values on an assumption that the linear decrease in $R_H$ with increasing $V_G$ in the $V_G$ range from 3.75 to 5.5 V corresponds to the increase in the sheet density of the accumulated electrons [$\Delta N_e = (1 / eR_H (V_G = x) - 1 / eR_H (V_G = 0)) \times t$, where $t$ is thickness of the FeSe channel (8.3 nm).]. Then we built a phase diagram (Fig. 4). $T_c$ vs. ($V_G$ or $R_H$ at 40 K) are also shown in Fig. S7 for comparison. At $8.1 \times 10^{13}$ cm$^{-2}$ ($V_G = +4$ V), onset $T_c$ of 8.6 K, which is close to that of bulk (16), was observed as seen in Fig. 3B. If we suppose that the whole channel layer is accumulated by electrons, the average electron density is estimated to be $9.8 \times 10^{19}$ cm$^{-3}$. This value is consistent with the native carrier concentration in bulk FeSe, the order of $10^{20}$ cm$^{-3}$ (23). With linear increase in $\Delta N_e$ up to $1.4 \times 10^{15}$ cm$^{-2}$ (the average carrier density of $1.7 \times 10^{21}$ cm$^{-3}$), the maximum onset $T_c$ of 35 K was observed. This result suggests that FeSe has a potential exhibiting such a high $T_c$ if such high-density carrier doping is possible. Actually, it is reported that external high pressures lead to this high-$T_c$ range superconductivity (27–37 K) (24, 25) without impurity doping, demonstrating its high-$T_c$'s potential.

The above observation suggests that applying a gate bias changes the structure and/or the electronic state of the FeSe epitaxial films. Thus, we would like to discuss the origin of this high-$T_c$ superconductivity. First, we discuss the possibility of a chemical reaction between the FeSe channel and the DEME-TFSI ionic liquid. Fig. 5A shows XRD patterns of the FeSe film dipped in DEME-TFSI without an applied gate bias. No impurity phase was detected, which is confirmed by the wide-range pattern from $2\theta =$ 10–80° in Fig. S8A, indicating that neither a chemical reaction between the FeSe film and DEME-TFSI nor a lattice parameter change occurs. We then examined the effect of applying $V_G = +5$ V for 2 h (Fig. 5B). In this case, we also did not observe a change in the XRD patterns (no impurity phase was detected, as confirmed by the wide-range pattern from $2\theta =$ 10–80° in Fig. S8B). These results indicate that the FeSe layer is stable against DEME-TFSI solution, and the ions in DEME-TFSI do not intercalate into the FeSe lattice [the interaction of amide ions and/or ammonia molecules with the FeSe lattice induces a distinct expansion of the $c$-axis length (26)]. Thus, it is plausible that the origin of this high-$T_c$ superconductivity is field-accumulated carrier doping. We speculate that an electronic transition similar to that under high pressures would be related to this high-$T_c$ superconductivity because their maximum $T_c$ values are similar (27–37 K for the high pressure cases) (24, 25).

In summary, we grew high-quality ~10-nm-thick FeSe epitaxial films on STO (001) substrates by MBE and confirmed their insulating electrical property. Using the high-quality thin film as a channel layer, we fabricated an EDLT to induce high-$T_c$ superconductivity. Upon applying $V_G$ = +5.0 V, an insulator–superconductor transition was induced with an onset $T_c$ = 24 K. The highest $T_c$ of 35 K was obtained by applying $V_G$ = +5.5 V. This $T_c$ value is significantly enhanced compared with the value of bulk FeSe ($T_c \sim 8$ K). Note that the origin of the superconductivity is not in the STO substrate because the $T_c$ of STO is as low as $0.2 - 0.4$ K even if an EDLT structure is employed (8, 27). Hall effect measurements suggest that the high-$T_c$ superconductivity comes from the highly accumulated electron carries in the FeSe channel surface. The relationship between $T_c$ and accumulated carrier density indicates that $T_c$ in FeSe channel increases monotonically to a breakdown voltage ($V_G$ > +5.5 V). The present study provides a way to investigate superconducting transitions even with an insulating parent phase without alteration/deterioration by impurity doping.

## Experiments

Heteroepitaxial FeSe thin films were grown on (001)-oriented STO single crystals by the MBE technique. To obtain an atomically flat substrate surface, we performed wet etching of as-received STO using a buffered HF solution, and then thermally annealed at 1050 °C just before film growth (28). The base pressure of the MBE growth chamber was <1 × 10$^{-7}$ Pa. We used two types of Knudsen cells (K-cells) to extract pure Fe and Se molecular beams: a high-temperature K-cell with a carbon heater for Fe (purity: 99.99%) and a normal K-cell with a tantalum heater for Se (purity: 99.999%). The temperatures of both K-cells were optimized to 1100 °C for Fe and 140 °C for Se using a beam flux monitor near the substrate. The optimized substrate temperature was 500 °C. The film thicknesses were ~10 and 110 nm, which were determined by least-squares fitting to a fringe pattern obtained by X-ray reflectivity spectroscopy using Cu K$\alpha_1$ monochromated by a Ge (220) crystal. The surfaces of the STO substrates and the FeSe films were observed by in situ RHEED and atomic force microscopy (AFM) in ambient atmosphere.

The structures of the films, such as the crystal orientation, were precisely examined by XRD (source: monochromatic Cu K$\alpha_1$) with an analyser crystal located in front of the detector. The $\omega$-coupled $2\theta$ scans in the out-of-plane XRD measurements provided the crystallographic orientation of the film normal to the substrate surface. The tilting angle of the crystallites was obtained by $2\theta$-fixed $\omega$ scans (out-of-plane XRC). The in-plane crystallographic orientation (i.e., orientation parallel to the substrate surface) was determined by $\phi$-coupled $2\theta_\chi$ scans. A $2\theta_\chi$-fixed $\phi$ scan also provided the rotational symmetry of the lattice/crystallites in the film plane. All of the axis relations of XRD can be found in ref. 29. The chemical compositions of the films (i.e., atomic ratio of Fe and Se) were determined with an electron probe microanalyser using wavelength-dispersive spectroscopy mode.

Approximately 10-nm-thick FeSe epitaxial films on the STO (001) substrates were used as the transport channel of the EDLT. The FeSe channel layer with a six-terminal Hall bar geometry (channel size: 500 μm long and 200 μm wide) and the Au pad electrodes were deposited using shadow masks. After bonding metal wires to the Au pads with In metal, the channel region was covered with a silica-glass cup (fixed with an epoxy adhesive) to add the ionic liquid. We used the ionic liquid DEME-TFSI as the medium for the gate insulator. The ionic liquid was poured into the silica-glass cup, and then a Pt coil was inserted into the ionic liquid to serve as the gate electrode. All of these device setup processes were performed without exposure to air. The device structure is shown in Fig. 2A.

Transfer curves (i.e., the $V_G$ dependence of the drain current $I_D$) were measured with a semiconductor parameter analyzer at 220 K. The temperature dependences of the resistivity and sheet resistance ($\rho$–$T$ and $R_s$–$T$), and transverse resistance ($R_{xy}$, i.e., Hall effect) at 40 K of the films and EDLTs were measured by the four-probe method under external magnetic fields of up to 9 T under an applied gate bias ($V_G$) from 0 to +5.5 V. $V_G$ was applied at 220 K because chemical reaction between FeSe layers and DEME-TFSI occurs at higher temperatures (30) and this temperature is well above the rubber phase-transition temperature of DEME-TFSI 190 K (31).

**Acknowledgments**

This work was supported by the Ministry of Education, Culture, Sports, Science and Technology (MEXT) through Element Strategy Initiative to Form Core Research Center. H. Hiramatsu was also supported by the Japan Society for the Promotion of Science (JSPS) Grant-in-Aid for Young Scientists (A) Grant Number 25709058, JSPS Grant-in-Aid for Scientific Research on Innovative Areas "Nano Informatics" (Grant Number 25106007), and Support for Tokyotech Advanced Research (STAR).

Electric Double Layers of Ionic Liquids. *Adv. Funct. Mater.* 19(7): 1046–1053.

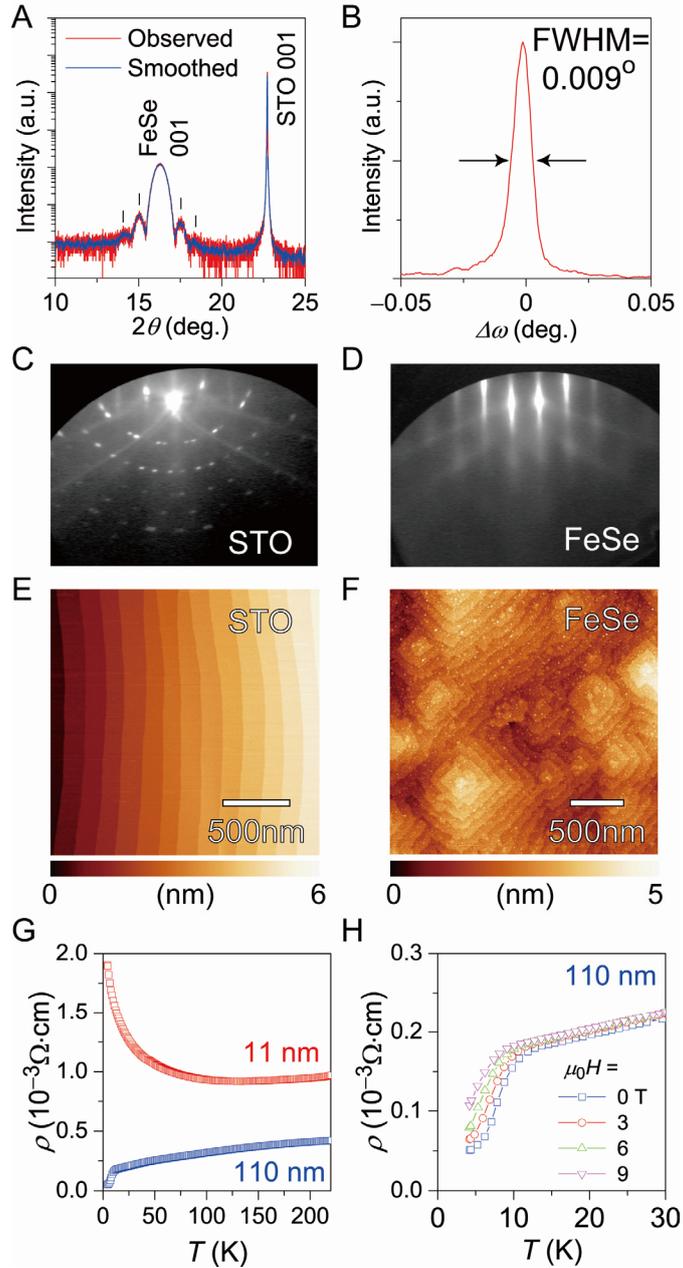

**Fig. 1.** Approximately 10-nm-thick FeSe films grown on STO (001) single-crystal substrates by MBE. (A) Out-of-plane XRD pattern. Vertical bars indicate Pendellösung interference fringes around FeSe 001 diffraction, originating from its flat surface. (B) Out-of-plane XRC of FeSe 001 diffraction. The full width at half maximum (FWHM) is 0.009°. RHEED patterns of the (C) STO substrate and (D) as-grown FeSe film surfaces. AFM images of the (E) STO substrate and (F) FeSe film surfaces. Horizontal bars

indicate the height scales for each image. (G) $\rho$–$T$ curves of the 11-nm-thick film compared with a 110-nm-thick film. (H) External magnetic field dependence of the $\rho$–$T$ curves of the 110-nm-thick sample, confirming that the film is a superconductor.

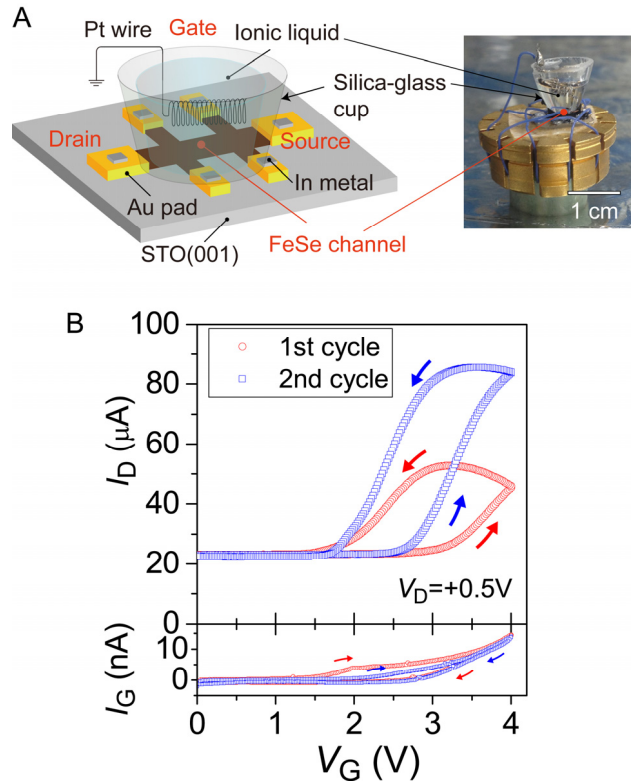

**Fig. 2.** (A) Device structure and photograph of the EDLT using the thin insulating FeSe layer as a channel. (B) Transfer characteristics (drain current $I_D$ versus gate bias $V_G$) of the FeSe EDLT under $V_D$ = +0.5 V at $T$ = 220 K cyclically measured for two loops. The arrows indicate the $V_G$-sweep directions starting from $V_G$ = 0 V. (Bottom) The leakage current ($I_G$) versus $V_G$ is also shown.

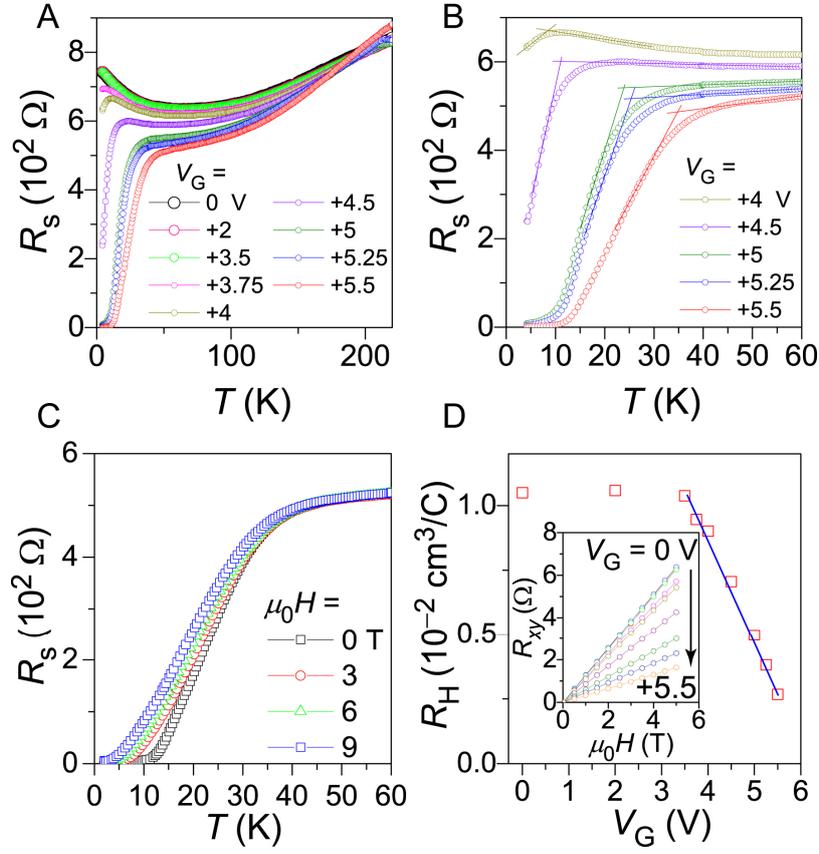

**Fig. 3.** Carrier transport properties of the FeSe EDLT under applied biases $V_G = 0 - +5.5$ V. (A) $R_s$–$T$ curves. (B) Enlarged image of A in the $T$ range $0 - 60$ K at $V_G = +4 - +5.5$ V. Solid lines are used to determine the onset $T_c$. (C) External magnetic field dependence of $R_s$–$T$ curves at $V_G = +5.5$ V. (D) Hall coefficients ($R_H$) at 40 K as a function of $V_G$. (Inset) Transverse Hall resistance ($R_{xy}$) at 40 K under external magnetic fields up to 5 T to obtain $R_H$.

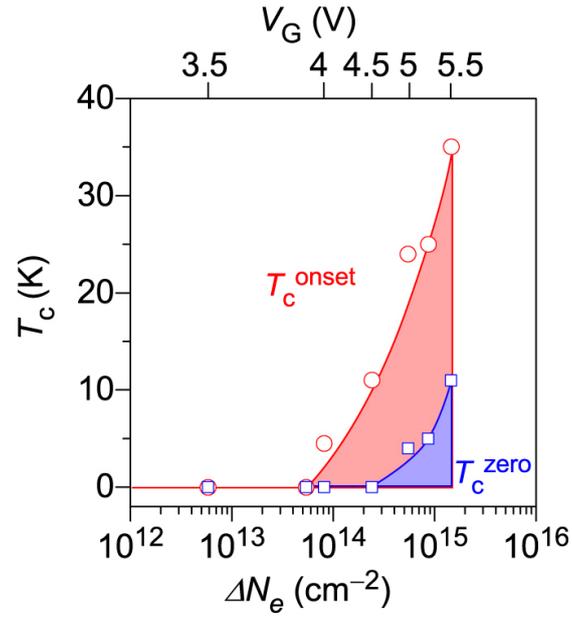

**Fig. 4.** Electronic phase diagram of the FeSe EDLT. Circles and squares show onset $T_c$ ($T_c^{onset}$) and zero resistivity temperature ($T_c^{zero}$), respectively. $\Delta N_e$ indicates the accumulated electron sheet carrier density (estimated) under $V_G$. The corresponding $V_G$ are shown in the upper horizontal axis.

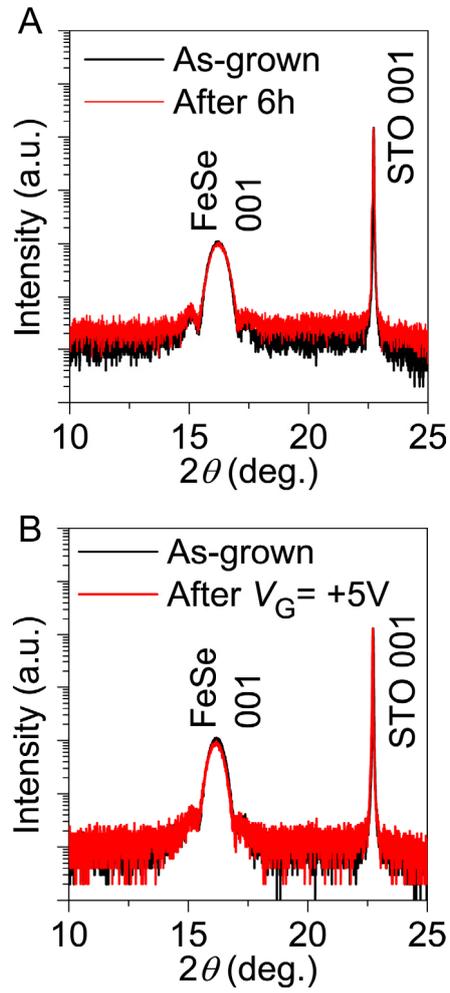

**Fig. 5.** XRD patterns of FeSe films dipped in DEME-TFSI. The intensity is normalized by that of STO 001 diffraction of the as-grown sample. (A) Dipped in DEME-TFSI for 6 h under a vacuum without applying a bias voltage. (B) Dipped in DEME-TFSI for 2 hours under a vacuum with $V_G$ = +5 V.

Supporting information for "Electric field-induced superconducting transition of insulating FeSe thin film at 35 K"

Kota Hanzawa,[1] Hikaru Sato,[1] Hidenori Hiramatsu,[1,2] Toshio Kamiya,[1,2] Hideo Hosono,[1,2]

[1]Materials and Structures Laboratory, Tokyo Institute of Technology, Mailbox R3-1, 4259 Nagatsuta-cho, Midori-ku, Yokohama 226-8503, Japan.

[2]Materials Research Center for Element Strategy, Tokyo Institute of Technology, Mailbox SE-6, 4259 Nagatsuta-cho, Midori-ku, Yokohama 226-8503, Japan.

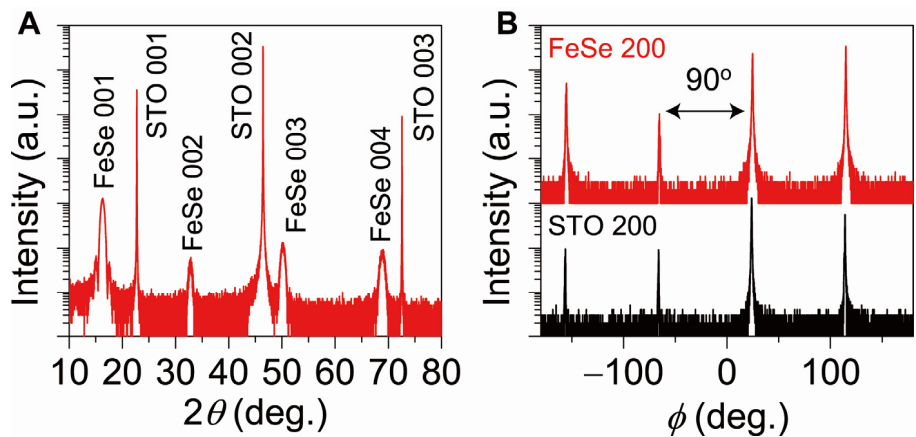

**Fig. S1.** XRD patterns of the epitaxial FeSe film grown on the STO (001) substrate. (A) Out-of-plane XRD pattern. (B) In-plane $\phi$ scans of STO 200 (Bottom) and FeSe 200 (Top) diffractions. These results indicate that the FeSe film grows heteroepitaxially on STO with the epitaxial relationship of [001] FeSe ∥ [001] STO for out-of-plane and [100] FeSe ∥ [100] STO for in-plane.

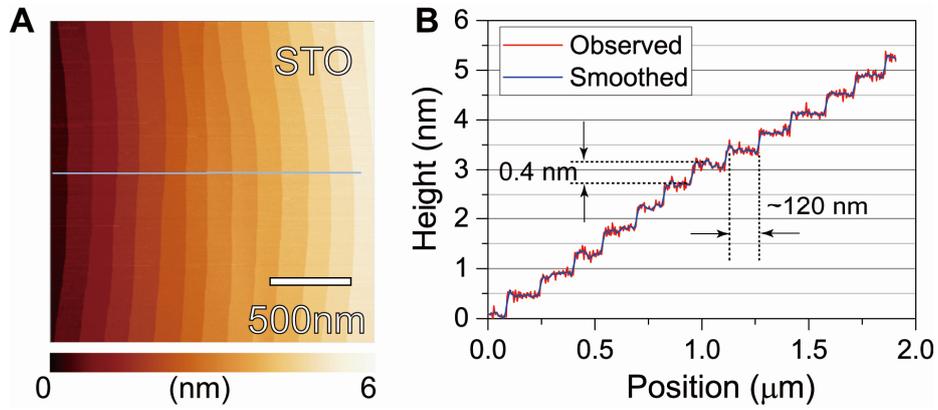

**Fig. S2.** (A) AFM image of the STO (001) substrate. (B) Cross-section along the horizontal line in A, indicating that the step height and terrace width are 0.4 and 120 nm, respectively. The step height is the same as the unit cell size of STO.

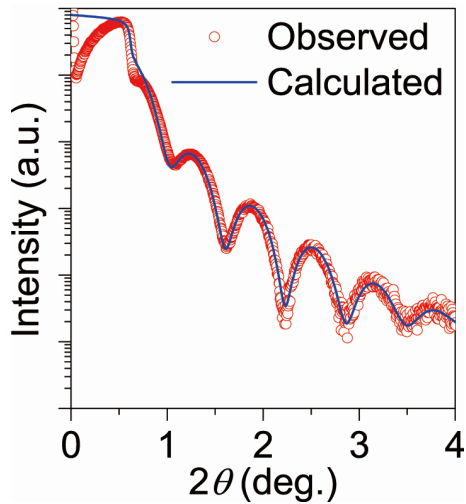

**Fig. S3.** X-ray reflectivity spectrum of the ~10-nm-thick FeSe epitaxial film. The thickness of this film is determined to be 12.4 nm from the fitting result.

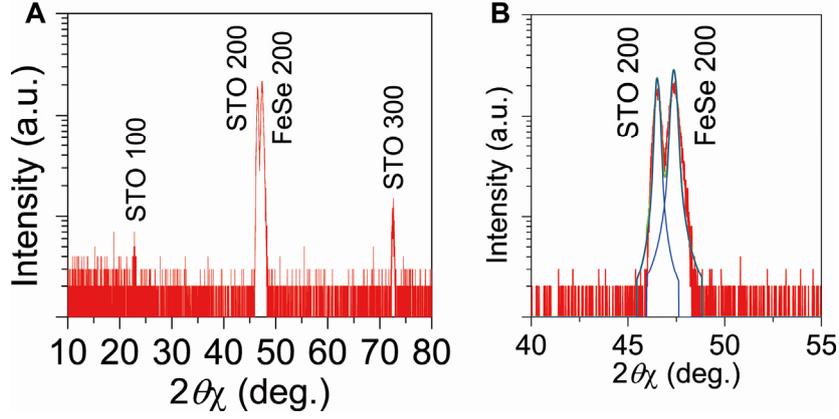

**Fig. S4.** In-plane XRD patterns of the 11-nm-thick FeSe film. (A) $\phi$-coupled $2\theta_\chi$ scan, and (B) enlarged pattern and deconvolution result to determine the in-plane lattice parameter. The calculated $a$-axis length is 0.3838 nm.

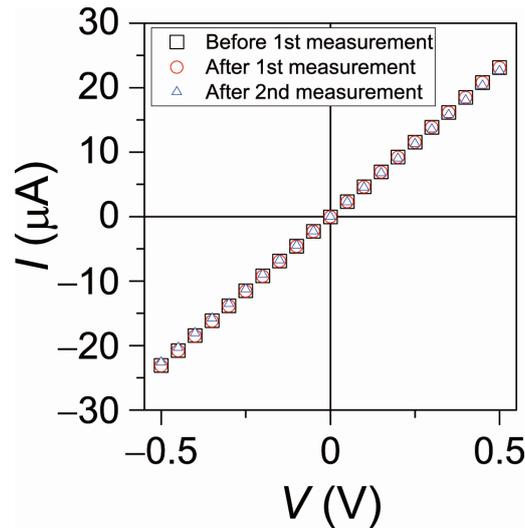

**Fig. S5.** $I$–$V$ characteristics of the FeSe EDLT at $V_G = 0$ measured at 220 K. Before (squares) and after (circles) the first cyclic measurement of the $I_D$–$V_G$ curve in Fig. 2B. The latter characteristic was unchanged before the 2nd cyclic measurement. After the 2nd cyclic measurement of the $I_D$–$V_G$ curve in Fig. 2B (triangles). These results confirm that there is no change in the channel layer.

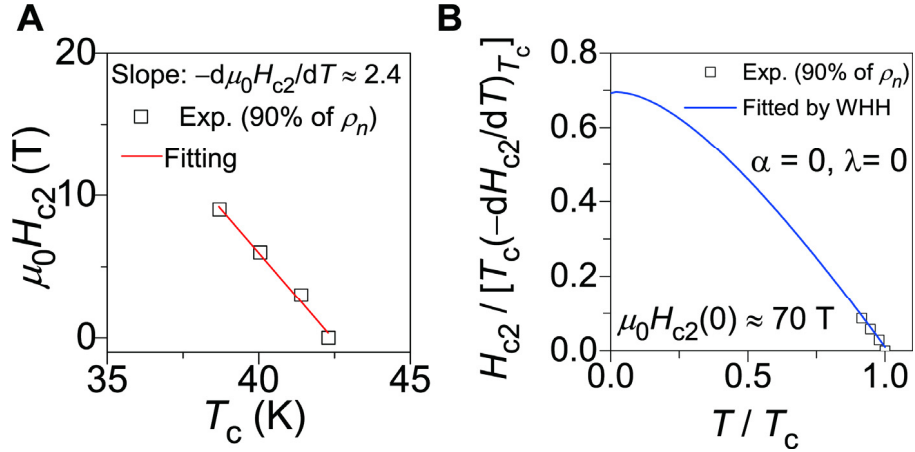

**Fig. S6.** Upper critical field ($\mu_0 H_{c2}$) as a function of temperature of FeSe EDLT under applying $V_G = +4.0$ V (Fig. 3D). (A) Least-square fitting result. (B) The estimated $\mu_0 H_{c2}$ at 0 K using Werthamer –Helfand–Hohenberg (WHH) theory (Ref. 20 in the main text) is 70 T. $\alpha$ and $\lambda$ demote Pauli spin paramagnetism and spin-orbit interaction parameters, respectively. Here, we select $\alpha = 0$ and $\lambda = 0$.

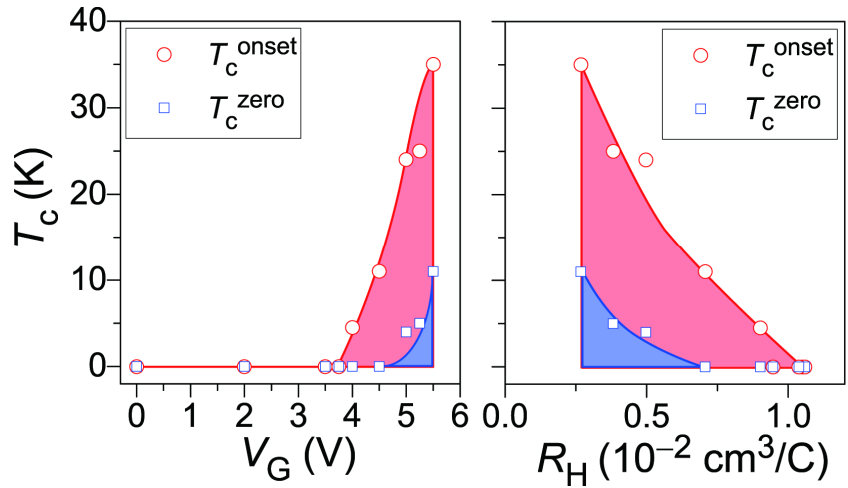

**Fig. S7.** Electronic phase diagrams of the FeSe EDLT. (Left) $T_c$ vs. $V_G$ and (Right) $T_c$ vs. $R_H$ at 40 K. Circles and squares show onset $T_c$ ($T_c^{onset}$) and zero resistivity temperature ($T_c^{zero}$), respectively. For a phase diagram ($T_c$ vs. estimated sheet carrier density), see Fig. 4 in the main text.

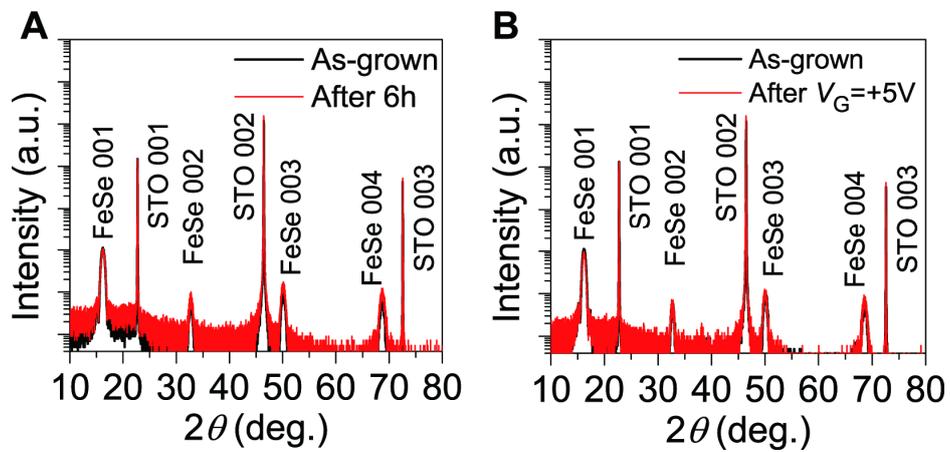

**Fig. S8.** XRD patterns of FeSe films dipped in DEME-TFSI. The intensity is normalized by that of STO 001 diffraction of the as-grown sample. (A) Dipped for 6 h under vacuum without an applied bias. (B) Dipped under vacuum with $V_G$ = +5 V.